\documentclass[AMA,STIX1COL]{WileyNJD-v2}

\articletype{Article Type}%

\received{26 April 2016}
\revised{6 June 2016}
\accepted{6 June 2016}

\usepackage{graphicx}
\usepackage{subcaption}
\usepackage{amsmath}
\usepackage{mathtools}
\usepackage[section]{placeins}
\usepackage{xcolor,colortbl}
\usepackage{multirow}
\definecolor{lightgray}{gray}{0.9}
\makeatletter
\newcommand\approxsim{\mathchoice
  {\@approxsim {\displaystyle}      {1ex} }
  {\@approxsim {\textstyle}         {1ex} }
  {\@approxsim {\scriptstyle}       {.7ex}}
  {\@approxsim {\scriptscriptstyle} {.5ex}}}
\newcommand\@approxsim[2]{%
  \mathrel{%
    \ooalign{%
      $\m@th#1\sim$\cr
      \hidewidth$\m@th#1.$\hidewidth\cr
      \hidewidth\raise #2 \hbox{$\m@th#1.$}\hidewidth\cr
    }%
  }%
}
\makeatother
\usepackage[colorinlistoftodos, textwidth=65mm, shadow]{todonotes}

\begin{document}

\title{Unlocking Insights: Enhanced Analysis of Covariance in General Factorial Designs through Multiple Contrast Tests under Variance Heteroscedasticity}

\author[1]{Matthias Becher*}
\author[2]{Ludwig A. Hothorn}
\author[1]{Frank Konietschke}

\authormark{BECHER \textsc{et al}}

\address[1]{\orgdiv{Institut für Biometrie und klinische Epidemiologie}, \orgname{Charité - Universitätsmedizin Berlin}, \orgaddress{\state{Berlin}, \country{Germany}}}

\address[2]{Im Grund 12, D-31867 Lauenau, Germany (retired from Leibniz University Hannover)}

\corres{*Corresponding author \email{matthias.becher@charite.de}}

\presentaddress{Matthias Becher, Institute of
Biometry and Clinical Epidemiology,
Charitéplatz 1, 10117 Berlin, Germany.}

\abstract[Summary]{A common goal in clinical trials is to conduct tests on estimated treatment effects adjusted for covariates such as age or sex. Analysis of Covariance (ANCOVA) is often used in these scenarios to test the global null hypothesis of no treatment effect using an $F$-test. However, in several samples, the $F$-test does not provide any information about individual null hypotheses and has strict assumptions such as variance homoscedasticity. We extend the method proposed by Konietschke et al. \cite{Konietschke2021} to a multiple contrast test procedure (MCTP), which allows us to test arbitrary linear hypotheses and provides information about the global- as well as the individual null hypotheses. Further, we can calculate compatible simultaneous confidence intervals for the individual effects. We derive a small sample size approximation of the distribution of the test statistic via a multivariate t-distribution. As an alternative, we introduce a Wild-bootstrap method. Extensive simulations show that our methods are applicable even when sample sizes are small. Their application is further illustrated within a real data example.}


\jnlcitation{\cname{%
\author{Becher M.},
\author{Hothorn L.}, and
\author{Konietschke F.}} (\cyear{2024}),
\ctitle{Unlocking Insights: Enhanced Analysis of Covariance in General Factorial Designs through Multiple Contrast Tests under Variance Heteroscedasticity}\cjournal{}, \cvol{2024}.}

\maketitle

\section{Introduction}\label{sec1}
A common goal in clinical and pre-clinical research is to estimate and test the impact of factor levels and their combinations on an outcome. Even in complete randomized designs, the outcome is often not just dependent on the factors of interest, but also on other variables such as age, gender, weight, etc. which are called covariates. They can distort the factor effects and thus bias the results and especially the treatment effect estimates. Therefore, the chosen model needs to adjust the treatment effect estimates for their impact, which is usually carried out in a so-called \textit{Analysis of Covariance} (ANCOVA) model. In statistical practice, especially in early and pre-clinical stages, researchers are often confronted with other data situations than (multivariate) normality and homogeneous variances; rather skewed data and especially variance heteroscedasticity are no rarity. The traditional $F$-test (which assumes homogeneity and normal errors) is thus often not applicable in these situations and therefore there is a need for statistical methods that neither postulate a specific data distribution nor homogeneous variances. \\
Different methods have been proposed to address the problem of variance heteroscedasticity in the general  ANOVA designs without covariates\cite{Pauly,Brunner}. More recently Konietschke et al. \cite{Konietschke2021} introduced a method for the general ANCOVA allowing for group-wise variance heteroscedasticity. In their paper they propose new test statistics, using methods of moments to get unbiased estimators of the variance components. They approximate the distribution of the test statistics with Box-type approximation methods.\\
One downside of these types of methods is that they can only be used to test the global null hypothesis $H_0$ of no effect difference between any groups. Finding significant differences in effects between specific groups of interest and computing corresponding simultaneous confidence intervals (SCIs) requires three steps. First, the global null hypothesis is tested. Then, multiple comparisons are tested to evaluate the individual hypotheses.  Lastly, the SCIs are computed. There are, however, some problems with this approach. The global null hypothesis may be rejected, but none of the individual and vice versa. In addition, even if an individual hypothesis is rejected the corresponding SCI may include the null, i.e. the value of no treatment effect. \\
To address these problems multiple contrast test procedures (MCTPs) were developed. These test procedures along with compatible SCIs were derived by Mukerjee et al. \cite{Mukerjee} and Bretz et al. \cite{Bretz}. Within the MCTP framework, multiple individual null hypotheses are tested with adequate test statistics. Since they are not necessarily independent, critical- and p-values are computed from their joint distribution taking their correlation into account. The global null hypothesis is rejected if any of the individual ones are rejected. This way the results of the individual null hypothesis tests are consonant and coherent with the result of the global null hypothesis test. MCTPs are highly flexible and can be used to test a variety of contrasts including all-pairs, many-to-one, trends, etc. As a result, they can be adapted depending on the research question. \\
Hothorn et al. \cite{Hothorn} have a proposed a method to apply the MCTP framework to general parametric models such as linear regression or ANCOVA. However, their method relies on distributional assumptions, at least asymptotically. In small sample sizes, the tests often behave liberal and over-reject the null hypotheses.    \\
In this article, we aim to close that gap and approximate the distributions of the tests for small sample sizes.  In addition, this approximation works under group-wise variance heteroscedasticity. Finally, we also introduce a wild-bootstrap-based method to estimate the distribution of the test statistic, when complete variance heteroscedasticity is present. The remainder of the paper is organized as follows: 
In Section \ref{sec2}, we introduce the statistical model as well as the effect and variance estimators. In Section \ref{sec3} we extend the introduced methods to the MCTP framework and present our approximations of the distribution of the test statistic. We will evaluate our new methods using a simulation study in Section \ref{sec4}. This is followed in Section \ref{sec5} by a data example using data from a toxicological animal study. Finally, we will discuss our findings in Section \ref{sec6}.

\section{Background}\label{sec2}
\subsection{Statistical model}
We consider a one-way layout involving $a$ groups of $n_i$ independent subjects each that have been (will be) observed under the respective treatments as well as under $M$ covariate conditions. Of major interest is estimating and testing treatment effect(s) adjusted for the impact of the covariates. Here and throughout, the assumption(s) on the variances of the data plays a major role in the inferential framework. The covariates and/or the treatments may induce heteroscedasticity group-wise (often in pre-clinical trials) or even subject-wise on an individual level (e.g., in observational trials). The common ANCOVA model assumes homogeneity and therefore does not constitute a suitable framework. Here and throughout, we consider a general ANCOVA model
\begin{eqnarray*} 
Y_{ij} = b_i + \sum\limits_{l=1}^{M}p_lM^{(l)}_{ij}+\epsilon_{ij},\; i = 1,\ldots,a;\; j = 1,\ldots,n_i, \nonumber
\end{eqnarray*}
where $b_i$ denotes the fixed and unknown effect of treatment $i$, and $p_l$ the regression parameter of the $l^{th}$ covariate. Two-way and higher-way layouts in means of general factorial designs are obtained by sub-indexing the index $i$. No less important, $\epsilon_{ij}$ denotes the error term with $E(\epsilon_{ij}) = 0$ and either group-wise variance $Var(\epsilon_{ij}) = \sigma_{i}^2 >0$ for $i=1,\ldots,a$, or subject-wise variances $\sigma_{ij}^2 >0$, for $i=1,\ldots,a; j=1,\ldots,n_i$, respectively. For ease of reading, we tackle both heteroscedastic models simultaneously and not separately. The test statistics proposed in the coming sections use variance estimators, which are consistent in either model, and therefore, the risk of confusing them is minimal. The total number of subjects is denoted by $N = \sum\limits_{i=1}^{a}n_{i}$. In matrix notation, the model can be written as
\begin{eqnarray} \label{model}
\mathbf{Y} = \mathbf{Xb} + \mathbf{Mp} + \mathbf{\epsilon}, \nonumber
\end{eqnarray} 
where $\mathbf{Y}$ denotes the $N \times 1$ response vector, $\mathbf{X} = \bigoplus^{a}_{i=1}\mathbf{1}_{n_i}$ the design matrix, $\mathbf{b} = (b_1,\ldots, b_a)'$ the vector of treatment effects, $\mathbf{M}$ the $N \times l$ matrix of the covariates, $\mathbf{p} = (p_1,\ldots, p_l)'$ the vector of regression parameters, and $\mathbf{\epsilon}$ the vector of the independent error terms with
\begin{eqnarray}\label{error}
    E(\mathbf{\epsilon})=\mathbf{0} \; \text{and} \; Var(\mathbf{\epsilon}) = \left\{
    \begin{array}{ll}
       \mathbf{\Sigma}_G = \bigoplus^{a}_{i=1}\sigma^{2}_i\mathbf{I}_{n_i},  & \text{(group-wise), \;\;or} \\
      \mathbf{\Sigma}_I = diag\left(\sigma^{2}_{11},\ldots,\sigma_{an_a}^2\right),   & \text{(subject-wise)}. 
    \end{array} \right.
\end{eqnarray}
The latter variance structure refers to \textit{complete} heteroscedasticity in which the random variables from each subject may have different variances and thus may come from different populations. Of major interest is making statistical inferences in estimating the unknown model parameters 
$\mathbf{b}$, $\mathbf{p}$, $\mathbf{\Sigma}_G$ and $\mathbf{\Sigma}_I$, as well as testing the null hypotheses
\begin{eqnarray}
    H_0^{(b)}: \mathbf{Cb} = \mathbf{0} \; \text{and} \; H_1^{(b)}: \mathbf{Cb} \neq \mathbf{0} \nonumber
\end{eqnarray}
formulated in terms of the treatment effects $\mathbf{b}$. Here, $\mathbf{C}$  denotes the hypothesis or contrast matrix, which we will discuss in more detail in the next subsection. 

\subsection{Hypotheses and contrast matrices} \label{sec: hypotheses}
In studies involving more than two randomized treatment levels of the factor, research questions are manifold and rarely restricted to testing the global null hypothesis $H_0^{(b)}: b_1=\ldots=b_a$ solely. If $H_0^{(b)}$ gets rejected by an appropriate test procedure (e.g., $F$-test), any of the $a$ groups differ. The arising question is ``Which ones?'', which then can only be answered using omnibus test procedures (post-hoc) and multiplicity adjustments. Hence, testing $q$ multiple local null hypotheses (alternatives) $H_0^{(\ell)}: \mathbf{c}_\ell' \mathbf{b}=0$ with a user-defined $q \times a$ contrast matrix $\mathbf{C}$ is of major interest. Here, $\mathbf{c}_\ell' = (c_{\ell 1},\ldots, c_{\ell a})$ denotes the $\ell$th row vector of $\mathbf{C}$, $\ell=1,\ldots,q$. For instance, many-to-one comparisons (so-called Dunnett-type contrasts \cite{dunnett1955multiple}) are performed using 
\begin{eqnarray}
     H_0: \left\{\begin{matrix}
   b_1 = b_2 \\
   b_1 = b_3 \\
   \vdots \\
   b_1 = b_a \\
        \end{matrix}\right.
        \Leftrightarrow
        H_0:\mathbf{Cb}= \begin{pmatrix}
                    -1 & 1 & 0 & 0 & \dots & 0\\
                    -1 & 0 & 1 & 0 & \dots & 0 \\
                    \vdots & \vdots & \vdots & \vdots & \vdots & \vdots \\
                    -1 & 0 & 0 & 0 & \dots & 1
                    \end{pmatrix}
                    \begin{pmatrix}
                    b_1 \\
                    b_2 \\
                    \vdots \\
                    b_a
                    \end{pmatrix} = \mathbf{0}, \nonumber
\end{eqnarray}
whereas all-pairwise comparisons (Tukey-type \cite{tukey1949comparing,tukey1953problem}) or comparisons to the mean (Grand-mean type \cite{Bretz}) are performed using 
\begin{eqnarray*}
&& H_0: \left\{\begin{matrix}
   b_1 = b_2 \\
   b_1 = b_3 \\
   \vdots \\
   b_1 = b_a \\
   b_2 = b_3 \\
   \vdots \\
   b_{a-1} = b_a \\
        \end{matrix}\right.
        \Leftrightarrow
        H_0:\mathbf{Cb}= \begin{pmatrix}
                    -1 & 1 & 0 & 0 & \dots & 0\\
                    -1 & 0 & 1 & 0 & \dots & 0 \\
                    \vdots & \vdots & \vdots & \vdots & \vdots & \vdots \\
                    -1 & 0 & 0 & 0 & \dots & 1 \\
                    0 & -1 & 1 & 0 & \dots & 0\\
                    0 & -1 & 0 & 1 & \dots & 0 \\
                    \vdots & \vdots & \vdots & \vdots & \vdots & \vdots \\
                    0 & 0 & 0 & \dots & -1 & 1
                    \end{pmatrix}
                    \begin{pmatrix}
                    b_1 \\
                    b_2 \\
                    \vdots \\
                    b_a
                    \end{pmatrix} = \mathbf{0}, \; \text{or} \\
&& H_0: \left\{\begin{matrix}
   b_1 = \Bar{b}_. \\
   b_2 = \Bar{b}_. \\
   \vdots \\
   b_a = \Bar{b}_. \\
        \end{matrix}\right.
        \Leftrightarrow
        H_0:\mathbf{Cb}= \begin{pmatrix}
                    1-1/a & -1/a & -1/a & -1/a & \dots & -1/a\\
                    -1/a & 1-1/a & -1/a & -1/a & \dots & -1/a \\
                    \vdots & \vdots & \vdots & \vdots & \vdots & \vdots \\
                    -1/a & -1/a & -1/a & -1/a & \dots & 1-1/a
                    \end{pmatrix}
                    \begin{pmatrix}
                    b_1 \\
                    b_2 \\
                    \vdots \\
                    b_a
                    \end{pmatrix} = \mathbf{0}, \Bar{b}_. = \frac{1}{a}\sum_{i=1}^a b_i,
\end{eqnarray*}
respectively. Which contrast to use depends on the research question of interest and cannot be recommended in a general way. Bretz et al. \cite{Bretz} provide an overview of different contrasts, which are also implemented in the \textit{R} function \textit{contrMat} within the \textit{multcomp}\cite{multcomp} package. Note that the Grand-mean type contrast matrix is the centering projection matrix $\mathbf{P}_a = \mathbf{I}_a - \mathbf{J}_a$ well known from linear model theory. 

\subsection{Effect and variance estimators} \label{sec: effect estimators}
To test the null hypotheses formulated above, the unknown model parameters $\mathbf{b}$ and $\mathbf{p}$ as well as the variances of their estimators must be estimated from the data. We use least squares for the former and obtain the well-known generalized least squares estimators (GLS)
\begin{eqnarray}
    \label{b_est}
    \mathbf{\widehat{b}}_c = \left(\mathbf{X'}\mathbf{\Sigma}_{c}^{-1}\mathbf{X}\right)^{-1}\mathbf{X'}\mathbf{\Sigma}_{c}^{-1}\left(\mathbf{Y} - \mathbf{M}\widehat{\mathbf{p}}_c\right) \;\; \text{and} \;\;
    \mathbf{\widehat{p}}_c = \left(\mathbf{M'Q}\mathbf{\Sigma}_{c}^{-1}\mathbf{M}\right)^{-1}\mathbf{M'Q}\mathbf{\Sigma}_{c}^{-1}\mathbf{Y},\; c \in \{G,I\},\nonumber
\end{eqnarray}
with the projection matrices $\mathbf{P = X(X'X)^{-1}X'}$, $\mathbf{Q} = \mathbf{I}_N - \mathbf{P}$, and covariance matrix $\mathbf{\Sigma}_{c}$ as defined in \eqref{error}. As usual, the ordinary least squares estimator (OLS) is obtained with $\mathbf{\Sigma}_{c} = \mathbf{I}_N$. Since  
$\mathbf{\Sigma}_{G}$ and $\mathbf{\Sigma}_{I}$ are unknown, they must be replaced by consistent estimators. For the ease of representation, we define the generating matrices
\begin{eqnarray}
    \label{AD}
    \mathbf{A}_c = \left(\mathbf{M'Q}\mathbf{\Sigma}_{c}^{-1}\mathbf{M}\right)^{-1}\mathbf{M'Q}\mathbf{\Sigma}_{c}^{-1}\; \; \text{and} \;\;
    \mathbf{D}_c = \left(\mathbf{X'}\mathbf{\Sigma}_{c}^{-1}\mathbf{X}\right)^{-1}\mathbf{X'}\mathbf{\Sigma}_{c}^{-1} - \left(\mathbf{X'}\mathbf{\Sigma}_{c}^{-1}\mathbf{X}\right)^{-1}\mathbf{X'}\mathbf{\Sigma}_{c}^{-1}\mathbf{MA}_c, \; c \in \{G,I\},
\end{eqnarray}
with the use of which the estimators $\mathbf{\hat{b}}_c$ and $\mathbf{\hat{p}}_c$ as well as their variances can be easily written as $\mathbf{\hat{b}}_c =\mathbf{D}_c\mathbf{Y}$,  $\mathbf{\hat{p}}_c = \mathbf{A}_c\mathbf{Y}$, and
\begin{eqnarray} \label{PsiXi}
\mathbf{\Psi}_c = Cov(\mathbf{\widehat{b}}_c) = \mathbf{D}_c\mathbf{\Sigma}_{c}\mathbf{D'}_c \;\; \text{and}\;\; \mathbf{\Xi}_c = Cov(\mathbf{\widehat{p}}_c) = \mathbf{A}_c\mathbf{\Sigma}_{c}\mathbf{A'}_c, \; c \in \{G,I\},
\end{eqnarray}
respectively. This representation is charming because the covariance matrices $\mathbf{\Psi}_c$ and $\mathbf{\Xi}_c$ can be written as matrix products (\textit{Sandwich})\cite{zeileis2006object} with $\mathbf{\Sigma}_{G}$ and $\mathbf{\Sigma}_{I}$ being the only unknown components. An unbiased estimator\cite{Konietschke2021,Cao} of $\mathbf{\Sigma}_{G}$ is given by 
\begin{eqnarray} \label{hatSigmaG}
    \widehat{\mathbf{\Sigma}}_G = \bigoplus_{i=1}^a \widehat{\sigma}_1^2 \mathbf{1}_{n_i}, \;\text{where} \; \widehat{\sigma}_i^2=\frac{\mathbf{Y'}_i\mathbf{Q}_i\mathbf{Y}_i}{n_i-1-rank(\mathbf{M}_i)}
\end{eqnarray}
is an unbiased estimator of $\sigma_i^2$. Here,$\mathbf{Y}_i$ is the $n_i\times 1$ vector of the response in group $i$, $\mathbf{Q}_i = \mathbf{I}_{n_i}-\mathbf{B}_i(\mathbf{B'}_i\mathbf{B}_i)^{-1}\mathbf{B'}_i$ is a projection matrix, $\mathbf{B}_i = (\mathbf{X}_i,\mathbf{M}_i)$, $\mathbf{X}_i = \mathbf{1}_{n_i}$, and $\mathbf{M}_i$ is the group specific matrix of covariates, respectively. Similarly, a heteroscedasticity consistent estimator\cite{white1980heteroskedasticity,Zimmermann} of $\mathbf{\Sigma}_I$ is given by
\begin{eqnarray} \label{boot_cov}
    \widehat{\mathbf{\Sigma}}_I=\bigoplus^a_{i = 1}\bigoplus^{n_i}_{j=1}\widehat{\epsilon}^2_{ij}.
\end{eqnarray}
Finally, we obtain consistent estimators $\mathbf{\widehat{\Psi}}_c$ and $\mathbf{\widehat{\Xi}}_c$ of $\mathbf{\Psi}_c$ and $\mathbf{\Xi}_G$ by replacing $\mathbf{\Sigma}_G$ and $\mathbf{\Sigma}_I$ with their estimators $\widehat{\mathbf{\Sigma}}_G$ and $\widehat{\mathbf{\Sigma}}_I$ in \eqref{PsiXi}, respectively.
For making inferences, information about the (asymptotic) joint distribution of the estimators is the only remaining missing information at this stage. It can be shown that for increasing (large) sample sizes, i.e., if $N \rightarrow \infty$ such that $N/n_i \rightarrow \lambda_i < \infty$, and under a few mild assumptions on the variances, the estimators follow multivariate normal distributions, i.e.,
\begin{eqnarray*}
    \sqrt{N}\left(\mathbf{\hat{b}}_c-\mathbf{b}\right) \approxsim N\left(\mathbf{0},N\mathbf{\Psi}_c\right), \; \text{and} \; 
    \sqrt{N}\left(\mathbf{\hat{p}}_c-\mathbf{p}\right) \approxsim N\left(\mathbf{0}, N\mathbf{\Xi}_c\right), 
\end{eqnarray*}
where $\mathbf{\Psi}_c$ and $\mathbf{\Xi}_c$ are as given in \eqref{PsiXi}, respectively. Next, different test statistics for testing the null hypotheses motivated above in Section~\ref{sec: hypotheses} will be introduced. 

\iffalse
We follow the approach proposed by Konietschke et al.\cite{Konietschke2021}. They estimate the variance components $\sigma_1^2,...,\sigma_a^2$ using the methods of moments on a group-specific level as they are model constants. They follow Cao et al.\cite{Cao} and estimate the using the group-wise sub-models. The estimator is given as
\begin{eqnarray}
    \label{variance_estimator}
    \hat{\sigma}_i^2=\frac{\mathbf{Y}'_i\mathbf{Q}_i\mathbf{Y}_i}{n_i-1-rank(\mathbf{M}_i)},i=1,...,a
\end{eqnarray}
$\mathbf{Q}_i = \mathbf{I}_{n_i}-\mathbf{B}_i(\mathbf{B}'_i\mathbf{B}_i)^{-1}\mathbf{B}'_i$ is a projection matrix, with $\mathbf{B}_i = (\mathbf{X}_i,\mathbf{M}_i)$. $\mathbf{X}_i = \mathbf{1}_{n_i}$ and $\mathbf{M}_i$ is the group specific covariate matrix.\\
Based on the unbiased variance estimator \eqref{variance_estimator} they derive unbiased estimators for $\mathbf{\Sigma,\;\Psi},\;\text{and}\;\mathbf{\Xi}$
\begin{eqnarray}\label{Psi}
    \mathbf{\hat{\Sigma}} = \bigoplus_{i=1}^a\hat{\sigma}_i^2\mathbf{I}_{n_i} \nonumber \\
    \mathbf{\hat{\Psi} = ND\hat{\Sigma}D'}  \\
    \mathbf{\hat{\Xi} = NA\hat{\Sigma}A'} \nonumber
\end{eqnarray}
\fi

\section{Multiple contrast test procedures}\label{sec3}
Since the pioneering works of Dunnett and Tukey \cite{dunnett1955multiple, tukey1949comparing}, and many others for making multiple comparisons, multiple contrast tests are well established tools in (bio)-statistical sciences. They have been developed for testing multiple null hypotheses formulated in different effects such as mean differences (or ratios thereof)\cite{Bretz,hasler2008multiple,hasler2014multiple, dilba2004simultaneous,dilba2007inferences}, proportions\cite{schaarschmidt2008approximate} or purely nonparametric relative effects\cite{konietschke2012rank, friedrich2017wild, rubarth2022ranking, gunawardana2019nonparametric}. In the following, we will revisit the general terminology for testing null hypotheses in ANCOVA models in line with Hothorn et al. \cite{Hothorn} and propose small sample size approximations of the distributions of the test statistics afterward. To test the individual null hypothesis $H_0^{(\ell)}:\mathbf{c}_\ell'\mathbf{b}=0$, consider the test statistic 
\begin{eqnarray} \label{Tlc}
 T_{\ell,c} =  \frac{\mathbf{c_\ell'\widehat{b}_c}}{\sqrt{\mathbf{c}_\ell' \mathbf{\widehat{\Psi}}_c \mathbf{c}_\ell}}, \; c \in \{G,I\}, \; \ell =1,\ldots,q.
\end{eqnarray}
Even though each test statistic $T_{\ell,c}$ follows a $N(0,1)$ distribution for large samples sizes, its finite and small sample size distribution is yet unknown. Furthermore, the test statistics $T_{\ell,c}$ and $T_{\ell',c}, \ell \not = \ell'$, are not necessarily independent. Ignoring their dependency by using the Bonferroni inequality within the inferential framework would result in a loss of power. We therefore collect them in the vector $\mathbf{T}_c=\left(T_{1,c},\ldots,T_{q,c}\right)'$ and use its joint distribution for the computation of critical- and p-values. Here, $\mathbf{T}_c$ follows, for large sample sizes, if $N\to \infty$ such that $\tfrac{N}{n_i}\leq N_0 < \infty$ (and further mild assumptions on variances being finite), a multivariate normal distribution with expectation $\mathbf{0}$ and correlation matrix 
\begin{eqnarray}\label{corMat}
    \mathbf{R}_c = diag\left(\mathbf{C}\mathbf{\Psi}_c\mathbf{C}'\right)^{-1/2} \left(\mathbf{C}\mathbf{\Psi}_c\mathbf{C}'\right) diag\left(\mathbf{C}\mathbf{\Psi}_c\mathbf{C}'\right)^{-1/2}.
\end{eqnarray}
For large sample sizes, the null hypothesis $H_0^{(\ell)}:\mathbf{c}_\ell'\mathbf{b}=0$ will be rejected at level $\alpha \in (0,1)$, if $|T_{\ell,c}| \geq z_{1-\alpha}( \mathbf{R}_c)$. A compatible $(1-\alpha)$ simultaneous confidence interval for the treatment effect $\mathbf{\delta}_\ell = \mathbf{c}_\ell'\mathbf{b}$ is obtained from 
\begin{eqnarray} \label{CIN}
 CI_\ell= \left[\mathbf{c}_{\ell}' \mathbf{\widehat{b}}_c \pm z_{1-\alpha}(\mathbf{R}_c)  \sqrt{\mathbf{c}_{\ell}^{'} \mathbf{\hat{\Psi}}_c \mathbf{c}_{\ell}}\right], \; \ell=1,\ldots,q,
\end{eqnarray}
where $z_{1-\alpha}(\mathbf{R}_c)$ denotes the two-sided $(1-\alpha)$-equicoordinate quantile of the $N(\mathbf{0},\mathbf{R}_c)$ distribution \cite{Hothorn}. For large sample sizes, the global null hypothesis $H_0: \mathbf{C}\mathbf{b}= \mathbf{0}$ will be rejected at level $\alpha$, if 
\begin{eqnarray}\label{T0N}
    T_{0,c} = \max\left\{|T_{1,c}|,\ldots,|T_{q,c}|\right\} \geq z_{1-\alpha}( \mathbf{R}_c).
\end{eqnarray}
For small sample sizes, however, the MCTP does not control the type-1 error rate, behaves liberal and over-rejects the null hypothesis, resulting in false positive conclusions. We therefore develop an approximation of the joint distribution of $\mathbf{T}_c$ for small sample sizes in the next subsections. We hereby differ between group-wise and complete heteroscedasticity and develop MCTPs for each model separately.

\subsection{A small size approximation of the distribution of $\mathbf{T}_G$}
\label{approx_section}
Group-wise heteroscedasticity is a common assumption in pre-clinical and clinical trials. Motivated by the Satterthwaite-Welch $t$-test and its generalization to MCTPs \cite{hasler2008multiple}, we aim to approximate the distribution of $\mathbf{T}_G$ by a central multivariate $t(\nu,\mathbf{R}_G,\mathbf{0})$-distribution with $\nu$ degrees of freedom and a correlation matrix $\mathbf{R}_G$. The correlation matrix is derived as shown in \eqref{corMat} based on the group-wise variance estimator we introduced in \eqref{hatSigmaG}. A straight forward computation of the degree of freedom $\nu$ is, however, not possible in the statistical model considered here: First, the multivariate $t$-distribution is, by definition, homoscedastic. Second, the variances of the contrasts $\mathbf{c}_\ell'\widehat{\mathbf{b}}_G$ differ in their degrees of heteroscedasticity. We therefore compute the degree of freedom of the distribution of each test statistic $T_{\ell, G}$ separately using a Box-type approximation, and select one of the $q$ candidates to approximate the joint distribution conservatively in a second step. Later, we compare the qualities of the different selections in extensive simulation studies in Section~\ref{sec4}. To begin with the former, we approximate the distribution of the estimated variance of $\mathbf{c}_\ell'\widehat{\mathbf{b}}_G$ by a scaled $\chi_{f_\ell}^2$ distribution,
\begin{eqnarray*}
    \mathbf{c}_{\ell}' \mathbf{\hat{\Psi}}_c \mathbf{c}_{\ell} \approx g_\ell \chi_{f_l}^2,
\end{eqnarray*}
such that the first two moments coincide and obtain

   \begin{eqnarray}
\label{df}
    \nu_\ell = \frac{diag(\mathbf{c^{'}_\ell\Psi c_\ell)^{2}}}{(\mathbf{k^{'}_\ell}\mathbf{X})^{2}\mathbf{D_{\hat{\sigma}}}^2\mathbf{\Omega}^{-1}}, \; \ell = 1,\ldots,q,
\end{eqnarray}
where, $\mathbf{k_\ell} = \mathbf{(c^{'}_\ell D)}^2$ denotes the $\ell$th row of the matrix $\mathbf{K} = \mathbf{(CD)}^2$, $\mathbf{D_{\hat{\sigma}}} = diag\left(\hat{\sigma}_1^2,\ldots, \hat{\sigma}_a^2\right)$ denote the diagonal matrix of the variance estimators $\widehat{\sigma}_i^2$ as given in \eqref{hatSigmaG} and  $\mathbf{\Omega^{-1}} = diag\left(\frac{1}{(n_1 - 1 - rank(\mathbf{M}_1))}, \ldots, \frac{1}{(n_a - 1 - rank(\mathbf{M}_a))}\right)'$ denote the diagonal matrix of their numerators (degrees of freedom) of each variance estimator separately. 
As potential candidates for the degree of freedom of the joint distribution, we choose the minimum, maximum, or mean (rounded to the nearest integer) of the candidate values $\nu_\ell$. The larger the degree of freedom the closer will be the joint distribution to the asymptotic $N(\mathbf{0}, \mathbf{R}_G)$ distribution, or, in other words, the larger $\nu$, the more liberal will be the approximation. The individual null hypothesis $H^{(\ell)}_0:\mathbf{c}_{\ell}^{'}\mathbf{b}=0$ will be rejected at multiple $\alpha$ level of significance, if $|T_\ell| \geq t_{1-a}(\nu, \mathbf{R}_G)$. Compatible $(1-\alpha)$-simultaneous confidence intervals are given by
\begin{eqnarray}\label{CIT}
    CI_\ell = \left[\mathbf{c}_{\ell}^{'} \mathbf{b} \pm t_{1-a}(\nu, \mathbf{R}_G)  \sqrt{\mathbf{c}_{\ell}^{'} \mathbf{\hat{\Psi}} \mathbf{c}_{\ell}}\right], \ell =1,\ldots,q.
\end{eqnarray} 
Finally, the global null hypothesis $H_0: \mathbf{Cb}=\mathbf{0}$ will be rejected at a significance level $\alpha$, if
\begin{eqnarray} \label{T0T}
    T_0 \geq t_{1-a}(\nu, \mathbf{R}_G) \nonumber
\end{eqnarray}
with $t_{1-a}(\nu, \mathbf{R})$ the ($1 - \alpha$)-equicoordinate quantile of the multivariate $t(\nu, \mathbf{R}_G, \mathbf{0})$ distribution. Note that the correlation matrix $\mathbf{R}_G$ is unknown in applications and must be estimated from the data. We recommend to replace $\mathbf{R}_G$ with the consistent estimator 
\begin{eqnarray*}
 \widehat{\mathbf{R}}_G = diag\left(\mathbf{C}\widehat{\mathbf{\Psi}}_G\mathbf{C}'\right)^{-1/2} \left(\mathbf{C}\widehat{\mathbf{\Psi}}_G\mathbf{C}'\right) diag\left(\mathbf{C}\widehat{\mathbf{\Psi}}_G\mathbf{C}'\right)^{-1/2}
\end{eqnarray*}
in \eqref{CIN},\eqref{T0N}, \eqref{CIT}, and \eqref{T0T} above, respectively.

\subsection{A small size approximation of the distribution of $\mathbf{T}_I$}
\label{bootstrap_section}
In the previous section, we introduced an approximation of the distribution of the maximum test statistic $T_{0,G}$ $ (\mathbf{T}_G)$ for small sample sizes under group-wise variance heteroscedasticity assumption, also known as the \textit{Behrens-Fisher} situation. In statistical practice, e.g., in observational studies, however, the assumption may be too strict and unrealistic. We therefore also investigate a multiple contrast test that is valid under complete heteroscedasticity. In this situation, approximating the distribution of the test statistic by a multivariate $t$-distribution results \textemdash if it is even possible \textemdash in very cumbersome computations of its degree of freedom. We, therefore, propose a Wild-Bootstrap resampling-based approximation as proposed by \cite{Konietschke2021, Zimmermann}. Let $W_{11},\ldots, W_{an_a}$ be independent and identically distributed random weights with $E(W_{11})=0$ and $Var(W_{11})=1$. We use random signs (Rademacher weights) throughout, i.e., $P(W_{11}=1)=P(W_{11}=-1)=1/2$, independently of the original data. The resampling variables are now computed by multiplying the random weights with the (scaled) residuals 
\begin{eqnarray}
    Y_{ij}^*=\widehat{\epsilon}_{ij}W_{ij}(1-p_{ij})^{-1/2}. \nonumber
\end{eqnarray} 
Here, $p_{ij}$ denotes the diagonal elements of the matrix $\mathbf{P}=\mathbf{BP}_B$, where $\mathbf{B}=\left(\mathbf{X},\mathbf{M}\right)$ and $\mathbf{P}_B = (\mathbf{B}'\mathbf{B})^{-1}\mathbf{B}'$ denotes a projection matrix, respectively. To approximate the distribution of $T_{0,I}$ for small sample sizes, consider the conditional (i.e., given the data $\mathbf{Y}$) distribution of the maximum test statistic
\begin{eqnarray} \label{boot_test}
  T_{0,I}^* = \max\left(|T_{1,I}^*|, ..., |T_{q,I}^*|\right), \; \text{where}\;\;   T_{\ell,I}^* = \frac{\mathbf{c}_\ell'\mathbf{\widehat{b}}_{I}^*}{\sqrt{\mathbf{c}_\ell'\mathbf{\widehat{\Psi}}_I^*\mathbf{c}_\ell}}, \; \ell=1,\ldots,q.
\end{eqnarray}
In comparison with the test statistic $T_{\ell,I}$ given in \eqref{Tlc}, $T_{\ell,I}^*$ is computed with the resampling variables $Y_{ij}^*$ instead of $Y_{ij}$, i.e., $\mathbf{\widehat{b}}_I^*$ denotes the OLS estimator as given in \eqref{b_est}, and $\mathbf{\hat{\Psi}}^*_I$ the upper left block matrix of the matrix $\mathbf{P_B}\mathbf{\hat{\Sigma}}_I^*\mathbf{P_B}'$ with $\mathbf{\hat{\Sigma}}_I^*$ denoting the  estimator as given in \eqref{boot_cov} computed with $Y_{ij}^*$ instead of $Y_{ij}$, respectively.
The individual null hypothesis $H_0^{(\ell)} : \mathbf{c}_{\ell}^{'}\mathbf{b}=0$ will be rejected at level $\alpha \in (0,1)$, if
\begin{eqnarray*}
    |T_{\ell,I}| \geq T_{1-\alpha}^\ast,
\end{eqnarray*}
where $T_{1-\alpha}^\ast$ denotes the $(1-\alpha)\cdot 100\%$ quantile of the distribution of $T_{0,I}^*$ as given in \eqref{boot_test}.
Compatible simultaneous confidence intervals for the treatment effects $\mathbf{\delta}_\ell=\mathbf{c}_\ell'\mathbf{b}$ are obtained from 
\begin{eqnarray*}
    CI_\ell^\ast = \left[\mathbf{c}_{\ell}^{'} \widehat{\mathbf{b}} \pm T_{1-\alpha}^\ast {\sqrt{\mathbf{c}_\ell'\mathbf{\widehat{\Psi}}^*_I\mathbf{c}_\ell}}\right].
\end{eqnarray*} 
Finally, the global null hypothesis $H_0: \mathbf{Cb}= \mathbf{0}$, will be rejected at level $\alpha$, if 
\begin{eqnarray} \label{T0star}
    T_{0,I} \geq T_{1-\alpha}^*.
\end{eqnarray}
The numerical computation of the critical values $T_{1-\alpha}^*$ (or p-values) is as follows:
\begin{enumerate}
    \item Given the data, compute the test statistic $T_{0,I}$ in \eqref{T0N}.
    \item Generate $N$ random weights $W_{ij}$ and compute the statistic $T_{0,I}^*$. Safe the value in $T_{0, I,1}$.
    \item Repeat the previous step a $n_{boot}$ times, e.g., $n_{boot}=10,000$, and obtain the values $T_{0,I,1},\ldots,T_{0,I,n_{boot}}$. 
    \item Estimate the critical value $T_{1-\alpha}^*$ by the empirical $(1-\alpha)\times 100\%$-quantile of the values $T_{0,I,1},\ldots,T_{0,I,n_{boot}}$.
\end{enumerate}

Note that the bootstrap approach is constructed in such a way that the correlation matrix of the multiple test statistics $T_{\ell, I}$ must not be estimated from the data. The information about the correlation is implicitly involved in the (resampling) distribution of $T_{0, I}^*$ (i.e., in $T_{1-\alpha}^*$). The approximation quality will be investigated in extensive simulation studies in the next section. 
\section{Simulation}\label{sec4}
\subsection{Setup}
All of the methods investigated are approximate. We therefore examine their behavior (type-I error rate control and power to detect selected alternatives) in small samples in an extensive simulation study; each with $nsim=nboot=10K$ simulation and bootstrap runs.
As a benchmark we use the general linear hypothesis multiple comparisons for parametric models (glht) proposed by Hothorn et al. \cite{Hothorn}, which we mentioned in chapter \ref{sec1}. We compare both the bootstrap method from \ref{bootstrap_section} and the approximation with the t-distribution as described in \ref{approx_section} to the benchmark. For the latter, we consider three possibilities (min, max, mean) to choose from the candidates for the degrees of freedom of the joint distribution \eqref{df}.\\
First, we examine the ability of the methods to accurately control the nominal type-1 error rate $\alpha = 5\%$. In the second step, we conduct further simulation studies to estimate the power of the methods to detect selected alternatives.
We would like to investigate whether the methods are applicable in a range of possible settings. Therefore, we generate data using different parameter combinations to emulate various scenarios. Data has been generated from

\begin{eqnarray}
    Y_{ik} &=& b_i+\mathbf{M}_{ik}^{'}\mathbf{p}+\epsilon_{ik}, i = 1,\dots,a; k = 1,\dots,n_i,\nonumber \\ 
    \epsilon_{ik} &=& \sigma_{ik} * \frac{Z_{ik} - E(Z_{ik})}{\sqrt{Var(Z_{ik}})}\nonumber
\end{eqnarray}
Here, $Z_{ik}$ are random values to base the error terms drawn from one of the following distributions: $N(0,1)$, $t_5$, $\chi^2_{12}$, or $Exp(1)$ to evaluate the methods in symmetric and tailed distributional scenarios 
with different numbers of groups ($a \in \{3,4,5\}$) and sample sizes. Throughout, we use m=4 covariates drawn from $N(7,1)$ and set $b_i=7$ (under the null hypothesis). The covariate effects are given by $\mathbf{p} = (0.2,1,1.5,2)'$.  In the case of variance homoscedasticity, we set $\sigma_1 = \ldots = \sigma_4 = \sigma_5 = 1$; otherwise in case of group-wise heteroscedasticity we choose $\sigma_1 \in \{2,4,6\}; \sigma_2 = 1.5; \sigma_3 = 1; \sigma_4 = 0.5; \sigma_5 = 0.75$, and for complete variance heteroscedasticity $\sigma_{ik} \sim U(0.5, 4)$, respectively. \\ We examine balanced and unbalanced designs with sample sizes $\mathbf{n} = (8,8,8,8,8) + i, \text{with} \; i \in \{0,2,4,\ldots,18\}$ in the balanced design and $\mathbf{n} = (8,10,13,17,20) + i \text{ or } \mathbf{n} = (20, 17, 13, 10, 8) + i \text{ for negative pairing (NP) or positive pairing (PP) respectively}, \text{with} \; i \in \{0,2,4,\ldots,18\}$. We investigate the impact of different contrasts ("Dunnett", "Tukey", and "GrandMean") on their type-I error rate control. \\
The implementation and simulation were done in R\cite{R} version 4.2.2. Computation has been performed on the HPC for Research cluster of the Berlin Institute of Health. The code for the simulation and the data example can be found in the following repository on the GitLab of the Charité (https://git-ext.charite.de/matthias.becher/mctp-ancova).
\subsection{Results}
Due to the many different settings, we will only present a few selected scenarios and discuss general findings in this section. The full results are provided in the supplementary material.\\

\begin{table*}[!htb]
\caption{Overview of the different parameter constellations and statistical models being simulated.\label{tab1}} 
\begin{tabular*}{\textwidth}{@{\extracolsep\fill}lllll@{}}\toprule 
\textbf{Setting Number} & \textbf{Number of Groups} & \textbf{Balanced/Unbalanced}  & \textbf{Variance Structure} \\ \midrule 
1 & 3 & Unbalanced NP& Group-wise heteroscedasticity \\ 
2 & 5 & Balanced & Complete heteroscedasticity \\ 
3 & 3 & Balanced & Homoscedasticity \\ 
4 & 4 & Unbalanced & Homoscedasticity \\ 
5 & 3 & Unbalanced & Complete heteroscedasticity \\ 
\bottomrule 
\end{tabular*}
\label{table:settings}
\end{table*}

\begin{figure}[!htb]
\includegraphics[width = \textwidth]{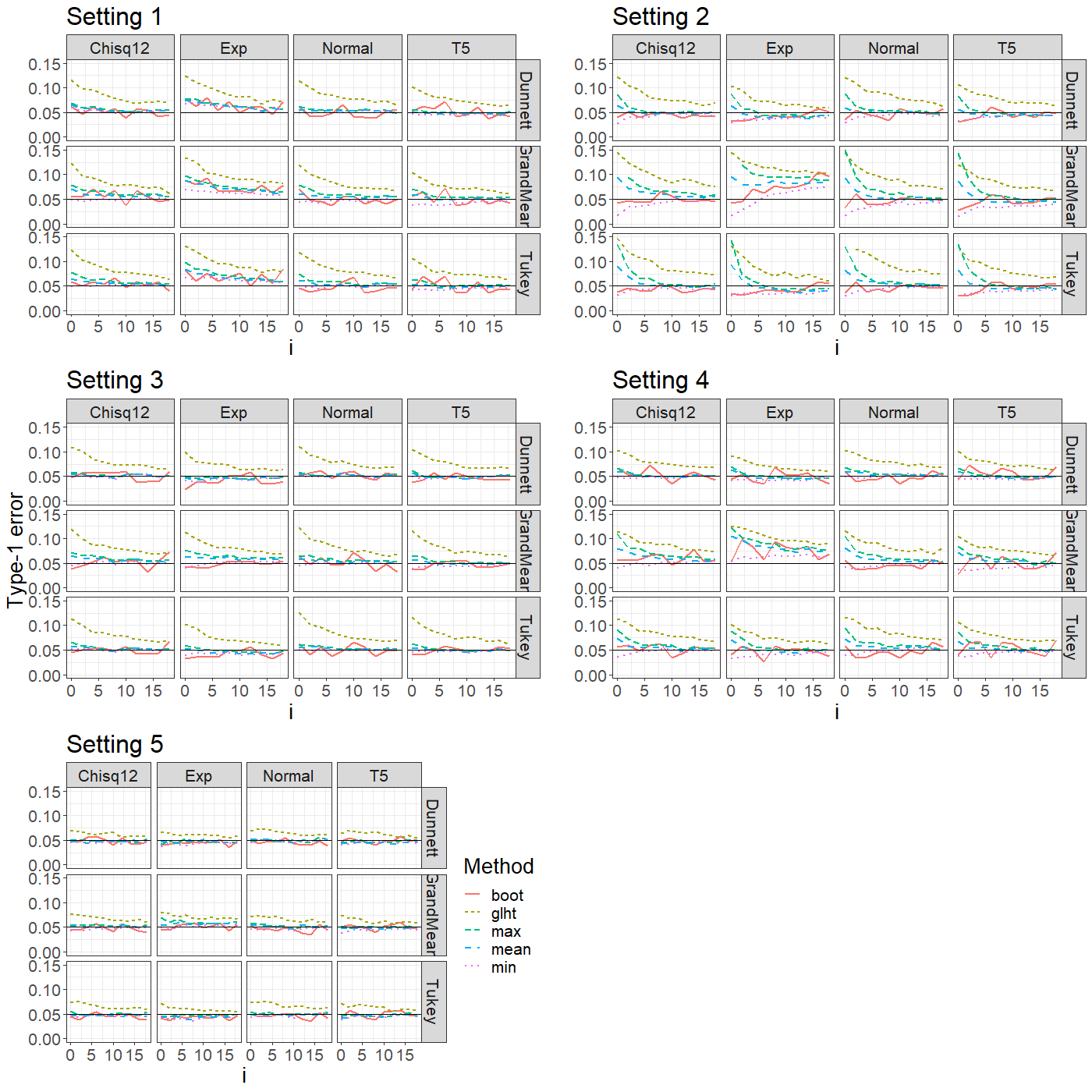}
\caption{Type-I error simulation results of the competing methods in the five settings given in Table~\ref{table:settings}.}
\label{fig:results}
\end{figure}

The different parameter constellations and models simulated are summarized in Table~\ref{table:settings}. For each setting, we show the results for all different distributions and contrasts by sample size. The results are shown in Figure \ref{fig:results}. \\
There are a few general observations we can make regardless of the setting. Firstly, we can see that glht is quite liberal in all settings and only converges towards the nominal type-1 error rate slowly with increasing sample sizes. Our methods outperform the competitors in all settings. Another observation we can make is that the bootstrap method appears to be slightly less stable than non-bootstrap methods. However, for very small sample sizes the bootstrap method performs best overall.\\
In setting 1 there are a few important points. As mentioned, we can see that glht is quite liberal, especially for small sample sizes, and converges to the nominal type-1 error rate with an increasing number of observations. In comparison, our methods generally work well even when sample sizes are small, even though mean and max tend to be liberal with very few observations. We can also see that the methods perform best with symmetrical distributions, while skewed data, especially the exponential, poses a bigger challenge. This is expected, but overall the performance is still quite good. \\

In Setting 2 in which contrasts of $a=5$ groups are tested, we can see that the methods behave more liberally than in situations with fewer groups when sample sizes are small. The bootstrap method generally outperforms the other methods for small sample sizes. As before, glht is liberal in all scenarios and never manages to reach the $5\%$ even with a higher sample size. In this setting, the multivariate t-approximation methods also tend to be liberal in small samples. While the mean and max methods are liberal, min is overly conservative. However, the type-I errors of the methods converge quickly towards the nominal $5\%$ level and show promising results when a moderate number of samples are available. The one exception to the overall good performance in Setting 2 is the case with the exponential distribution and the GrandMean contrast. There the methods are converging towards around $8\%$. They tend to be accurate with large sample sizes $(n=100)$. That being said, this specific setting is a somewhat extreme scenario and the results are expected. The performance overall in Setting 2, especially from the bootstrap method is very good.\\ 
Setting 3 is probably the "easiest" setting. As expected all of our methods perform well across the board, although the bootstrap method is a bit less stable. Nevertheless, there are no major deviations and it manages to control the type-1 error rate well overall. In contrast, the competitor again is liberal for small sample sizes and only slowly converges towards the $5\%$ level of significance. \\
Settings 4 and 5 again show that among the 3 parameters we are varying here, the number of groups has the strongest impact on the performances of the methods. Both settings are unbalanced and Setting 5 has complete variance heteroscedasticity compared to homoscedastic variance in Setting 4, which should be more challenging. However, with 3 groups in Setting 5 being involved, all of our methods control the nominal type-1 error rate in all scenarios for almost all sample sizes. In contrast, with 4 groups in setting 4, the performance suffers especially when only a few samples are available. Similar to Setting 2 the bootstrap method performs best at very small sample sizes. \\
Overall, some general observations can be made across all the results. First, the choice of the research question and as a result the choice of the contrast has a sizeable effect on the performance of the methods. Since the GrandMean contrast induces a negative correlation among the test statistics, it is more demanding, particularly when combined with a skewed distribution like the exponential. This leads to the next observation. In general, symmetric distributions are handled a bit better. However, even for skewed distributions the performance of our methods, and in particular the bootstrap method, is still good, even with only a few or a moderate number of samples. As one would expect the performance gets worse with an increasing number of groups being compared. Somewhat counter-intuitively the performance is mostly better in the unbalanced design. This is likely due to the simulation setup where the unbalanced design has an overall larger sample size. Consequently, the total number of samples is more important than some imbalance. We also investigated settings where $N$ was constant between balanced and unbalanced scenarios (Figure \ref{fig:constantN} in the appendix), but there was no major difference between the two.  Results were slightly better for positive pairing compared to negative pairing, especially for small sample sizes. This is expected as negative pairing combines the problems of small sample size and large variance, whereas in positive pairing one is offset by the other. For a comparison we refer to Figure \ref{fig:npvspp} in the appendix. Interestingly, group-wise heteroscedasticity seems to be a bit more challenging for the bootstrap method, compared to complete heteroscedasticity, though it still generally performs well even in those scenarios.
\subsection{Power simulations}
\begin{figure}[!htb]
\includegraphics[width = \textwidth]{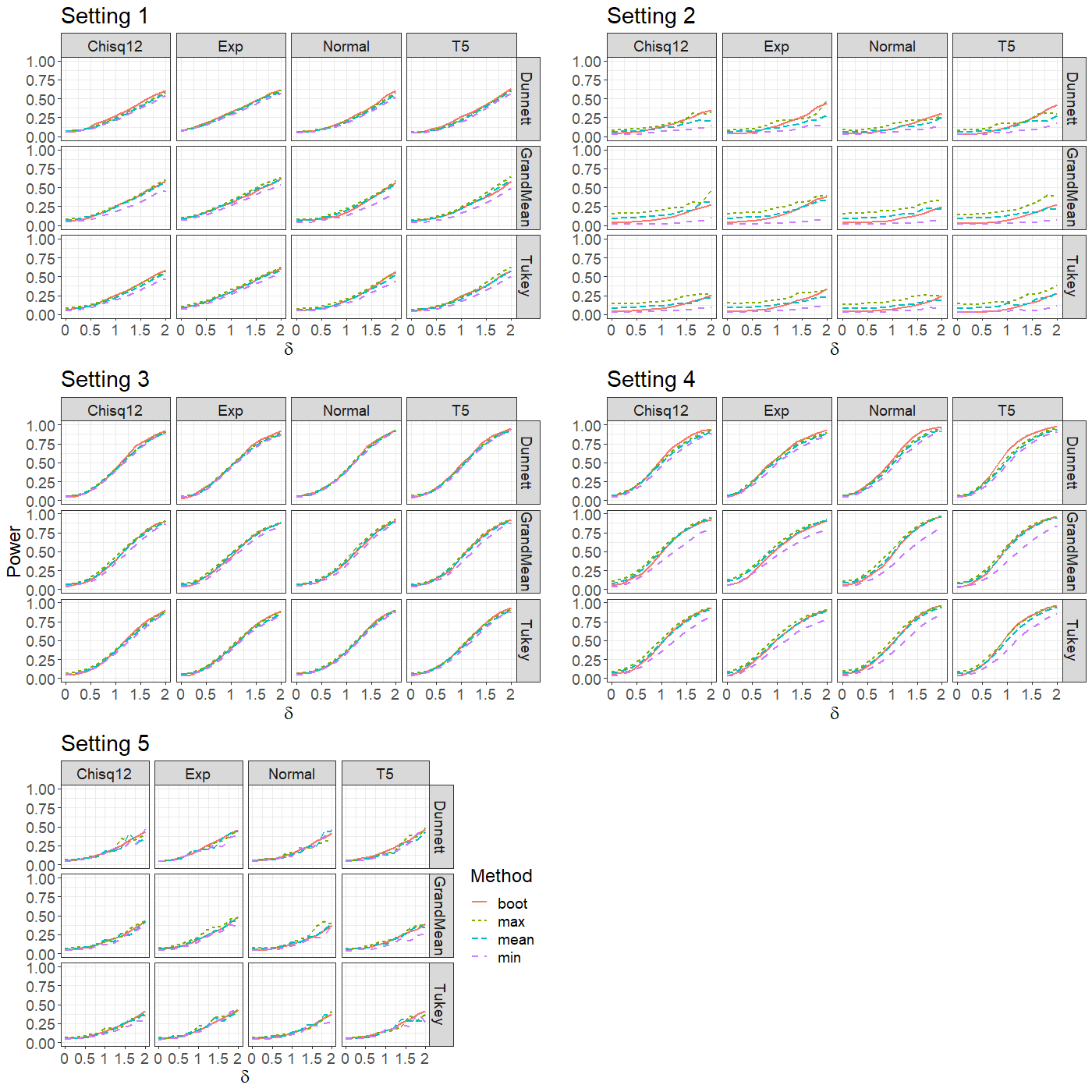}
\caption{Power analysis results for the 5 settings with alternative 1}
\label{fig:power1}
\end{figure}

In the previous section, we investigated the ability of our methods to control the type-1 error rate. In practice, we are also interested in the statistical power of the tests. To that end, we conduct an extensive power simulation study based on the simulation setup described earlier. Of course, there is a wide range of possible alternatives to test. We investigate two possible alternatives:
\begin{eqnarray*}
    \text{Alternative 1:} \;\;\mathbf{b}' = \left(7\mathbf{1}_{a-1}^{'} - \delta, 7\mathbf{1}_{a-1}^{'}\right)   \text{ and }
    \text{Alternative 2:} \;\;\mathbf{b}' = \left(7\mathbf{1}_{a-2}^{'} - \delta, 7\mathbf{1}_{a-2}^{'} + \delta, 7\mathbf{1}_{a-2}^{'}\right). 
\end{eqnarray*}
Here $\delta \in \{0, 0.2, \ldots, 2\}$ is the effect we want to detect. It is the same across all settings. We only estimate the power for the smallest sample sizes $\mathbf{n} = (8,8,8,8,8)$ and $\mathbf{n} = (8,10,13,17,20)$ in the balanced and unbalanced designs respectively. As we have seen the competitor is quite liberal in virtually all scenarios at that sample size. We therefore exclude it from the power analysis. \\
We present the results for the same settings we showed for the type-1 error simulation. The results for Alternative 1 are shown in Figure \ref{fig:power1}. We can see that the most important impact on the power is the variance. In Settings 3 and 4 with homoscedastic variances, the power rises much faster with an increasing effect $\delta$ compared to the group-wise heteroscedasticity (Setting 1) which in turn has a higher power compared to complete variance heteroscedasticity (settings 2 and 5). \\
Overall, the methods perform comparably across the 5 settings. The min-method tends to have slightly lower power, especially in the settings with more than 3 groups (Settings 2 and 4). This is in line with the previous results where it was the most conservative. \\
As before, positive pairing leads to better results compared to negative pairing, which can be seen in Figure \ref{fig:npvspp_power} in the appendix. Additionally, in the case of group-wise heteroscedasticity the variance of the first group $\sigma_1$ had a noticeable difference on the power, with a larger $\sigma_1$ and therefore a larger difference in variance between the groups, leading to lower power. \\
In the situations considered here, the choice of contrast seems to have a significant impact on the power of the methods; this is different in other alternatives. Furthermore, since all error distributions are standardized, they seem to impact the power only on a minor level. This is in contrast to the results we have seen in the type-1 error simulation. \\
The results for the same settings with alternative 2 are shown in Figure \ref{fig:power2} in the appendix. Overall, the results for Alternative 2 are similar to Alternative 1. The power generally is higher which we would expect, as there are two groups with effects compared to one. The most important factor influencing the power is still the variance structure. \\
We can also observe again, that the min-method tends to have the lowest power of our proposed methods.
\section{Data Example}\label{sec5}
To illustrate the need for the methods and their applications, we consider a subproject of a toxicological study on pyridine (number C55301B) from the US National Toxicology Program (https://manticore.niehs.nih.gov/cebssearch/test\_article/110-86-1; assessed November 2020). In this animal study, the researchers investigated the effect of pyridine on various clinical chemistry parameters. As the outcome of interest, we chose blood urea nitrate (BUN) measured in mg/dl. The study included N = 120 rats (60 male and 60 female) randomly assigned to one of six treatment groups, six dose levels of pyridine (0, 50, 100, 250, 500, 1000). BUN was measured twice, once at baseline and once after 90 days. BUN after 90 days is the outcome of interest. As covariates, we include BUN at baseline as well as the change in body weight during the trial. This trial can then be analyzed as a two-way factorial design with the factors dose (6 levels), gender (2 levels), and 2 covariates. Figure 3 shows a boxplot of the outcomes for each combination of the factor levels. 
\begin{figure}
\centering
\includegraphics[width = \textwidth]{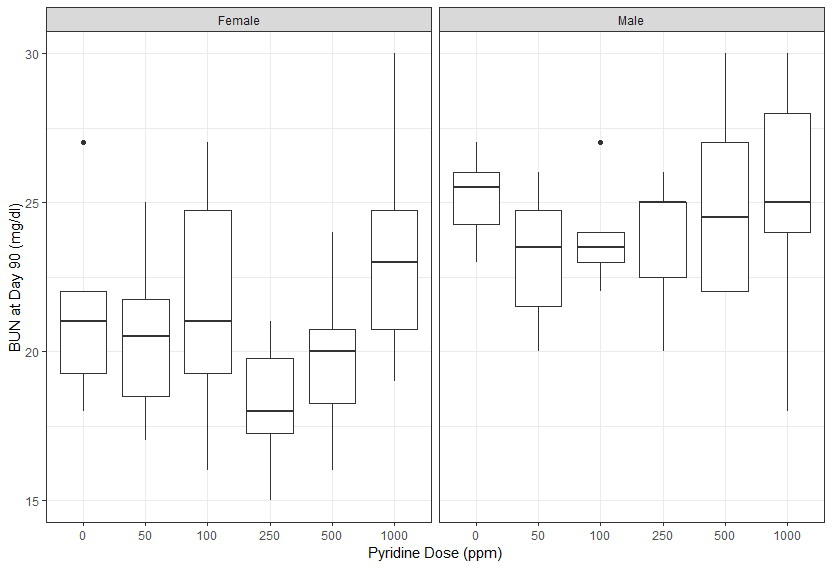}
\caption{BUN at day 90 stratified by factors sex and dose}
\label{fig:combined}
\end{figure}
We can see a difference in the outcome between male and female rats but no obvious relationship between dose levels and the outcome. Based on the boxplot no distributional assumptions can be made. In addition, the empirical variances between the different factor combinations are quite different, so we cannot assume variance homogeneity. \\
The trial can be statistically modeled using a general ANCOVA design 
\begin{eqnarray*}
    Y_{ijk} = b_{ij} + \sum_{l = 1}^2p_lM_{ijk}^{(l)} + \epsilon_{ijk}, i = 1,2; j=1,...,6; k = 1,...,n_i
\end{eqnarray*}
where $Y_{ijk}$ is BUN on day 90, $b_{ij}$ are the factor-wise treatment effects, $M_{ijk}^{(1)}$ and $M_{ijk}^{(2)}$ are BUN at baseline and change in body weight with $p_1$ and $p_2$ the corresponding coefficients. Their OLS estimates are $\hat{p}_1 = -0.036$ and $\hat{p}_2 = -0.009$, respectively, i.e., none of them significantly impacts the response. \\
We now use our method to test for group-wise differences for either the individual factors or the combination of the two. Testing the null hypotheses of no gender effect, no dose effect and no interaction between dose and gender with global testing procedures, e.g., with the ones proposed by \cite{Konietschke2021}, yields global significant dose and gender effects at $5\% $ level of significance. Which dose levels differ from each other is unknown but of paramount importance for the researcher. To illustrate the benefit of analyzing using MCTPs, we use the same contrasts as being used by the ANCOVA and show the results for testing the global null hypotheses in Table~\ref{global_results}. We see that all methods detect a significant effect on the individual factors. The data do not provide the evidence to reject the null hypothesis of no interaction effect.  Since the sample sizes are very small $(n \leq 10$ in every cell), we tested the interaction effect with the bootstrap method only. 
\begin{table}[!htb]
\caption{Global test results for the individual and combined effects for each our different methods \label{global_results}}
\centering
\begin{tabular}{rllrrlrr}
  \hline
  Method & Effect & Test statistic & p-value &  \\ 
  \hline
  \rowcolor{gray!10}
      & Dose & 3.525 & 0.015 & * \\ 
    \rowcolor{gray!10}
    \multirow{-2}{*}{min}
   & Sex & 4.591 & 0.000 & * \\ 
   & Dose & 3.525 & 0.009 & * \\ 
        \multirow{-2}{*}{mean}
   & Sex & 4.591 & 0.000 & * \\ 
  \rowcolor{gray!10}
   & Dose & 3.525 & 0.008 & * \\
  \rowcolor{gray!10}
    \multirow{-2}{*}{max} 
   & Sex & 4.591 & 0.000 & * \\
   & Dose & 3.544 & 0.013 & * \\ 
   & Sex & 4.291 & 0.000 & * \\ 
        \multirow{-3}{*}{boot} 
   & Dose $\times$ Sex & 5.231 & 0.139 &  \\ 
   \hline
\end{tabular}
\end{table}
As already mentioned, one of the advantages of the MCTP framework is the ability to also test the individual hypotheses. Based on the global results we know that there is a significant effect of the dose levels. However, we do not know which levels deviate from the average. Using MCTP we can look at the results for the individual hypotheses. Table \ref{individual_results} shows the results for individual dose levels calculated using the bootstrap method. This allows us to see not only that there is a difference between dose levels, but also which levels differ from the overall mean. In this case, dose levels 250 and 1000 both differ significantly from the overall mean at $5\%$ level of significance.  
\begin{table}[!htb]
\caption{Individual results for dose level estimated by the bootstrap method  \label{individual_results}}
\centering
\begin{tabular}{rlrrrrrl}
  \hline
 & Contrast & Effect & CI\_Lower & CI\_Upper & Test statistic & p-value  &  \\ 
  \hline
  \rowcolor{gray!10}
1 & 0 & 0.27 & -1.20 & 1.74 & 0.55 & 0.99 &  \\ 
  2 & 50 & -1.07 & -2.59 & 0.46 & 2.07 & 0.28 &  \\ 
  \rowcolor{gray!10}
  3 & 100 & -0.07 & -1.75 & 1.62 & 0.11 & 1.00 &  \\ 
  4 & 250 & -1.70 & -3.11 & -0.28 & 3.54 & 0.01 & * \\ 
  \rowcolor{gray!10}
  5 & 500 & 0.20 & -1.58 & 1.99 & 0.34 & 1.00 &  \\ 
  6 & 1000 & 2.77 & 0.42 & 5.12 & 3.49 & 0.01 & * \\ 
   \hline
\end{tabular}
\end{table}
\section{Discussion}\label{sec6}
ANCOVA is one of the most widely used methods in clinical and non-clinical research to estimate and test null hypotheses concerning adjusted treatment effects. However, common ANCOVA models have several important limitations. Firstly, as parametric models, they rely on strict distributional assumptions such as normality and variance homoscedasticity. Secondly, the commonly used ANCOVA $F$-test is limited to assessing the global null hypothesis of no difference in effects between groups only.

Konietschke et al. \cite{Konietschke2021} addressed the first issue by proposing an alternative approach for the analysis of covariance in general factorial designs using Box-type approximation methods, which are also valid under variance heteroscedasticity.

In this paper, we extended their approach to the MCTP framework, thereby addressing the second shortcoming. This extension allows us to test not only the global null hypothesis but also individual null hypotheses and construct consistent simultaneous confidence intervals. To achieve this, we derived a small sample size approximation for the distribution of the test statistics via a multivariate t-distribution, which does not necessitate variance homoscedasticity but is also applicable for group-wise variance heteroscedasticity. Additionally, we proposed an alternative method using a wild-bootstrap approach to empirically approximate the distribution of the test statistic through resampling. This method further relaxes the model assumptions and is also valid in the presence of complete variance heteroscedasticity in the data.

To evaluate the ability of our methods to control the type-1 error rate, we conducted a simulation with various settings. As a point of comparison, we utilized an approach by Hothorn et al. \cite{Hothorn}, which is based on general parametric models.

Our simulations demonstrated that our methods outperformed the competitor in all scenarios, particularly in small sample sizes. In complex settings with numerous groups and very few samples, only the bootstrap method outperformed the multivariate t-approximations. However, with increasing sample sizes, the type-I error rates of the methods quickly converged towards the nominal level, demonstrating their applicability even with a moderate number of samples.

Furthermore, we conducted a power analysis, which revealed that the degree of variance heteroscedasticity primarily impacts the power of the procedures. The procedures attained the highest/lowest power under homoscedasticity/complete heteroscedasticity, respectively. Except a few scenarios in which the min-method exhibited significantly lower power compared to the other methods, all methods displayed similar power.

Consequently, the general recommendation would be to employ the bootstrap method when faced with few samples or challenging scenarios such as skewed data or a large number of groups. The downside of the bootstrap method is its computational cost compared to the other methods. However, if a moderate or large number of observations ($n_i>15$) is available per group, then the approximate methods also consistently perform well. All of the methods discussed are mean-based and thus, only applicable in metric models. MCTP for nonparametric ANCOVA models, e.g., probabilistic index models \cite{de2015regression,thas2012probabilistic}, will be part of future research.
%\backmatter

\section*{Acknowledgments}
The work was supported by the Deutsche Forschungsgemeinschaft
grant number DFG KO 4680/4-1.

\subsection*{Financial disclosure}

None reported.

\subsection*{Conflict of interest}

The authors declare no potential conflict of interests.

\nocite{*}% Show all bib entries - both cited and uncited; comment this line to view only cited bib entries;
\bibliography{wileyNJD-AMA}%

\appendix
\section{Balanced vs Unbalanced with constant $N$ \label{app1}}
\begin{figure}[!htb]
\includegraphics[width = 160mm]{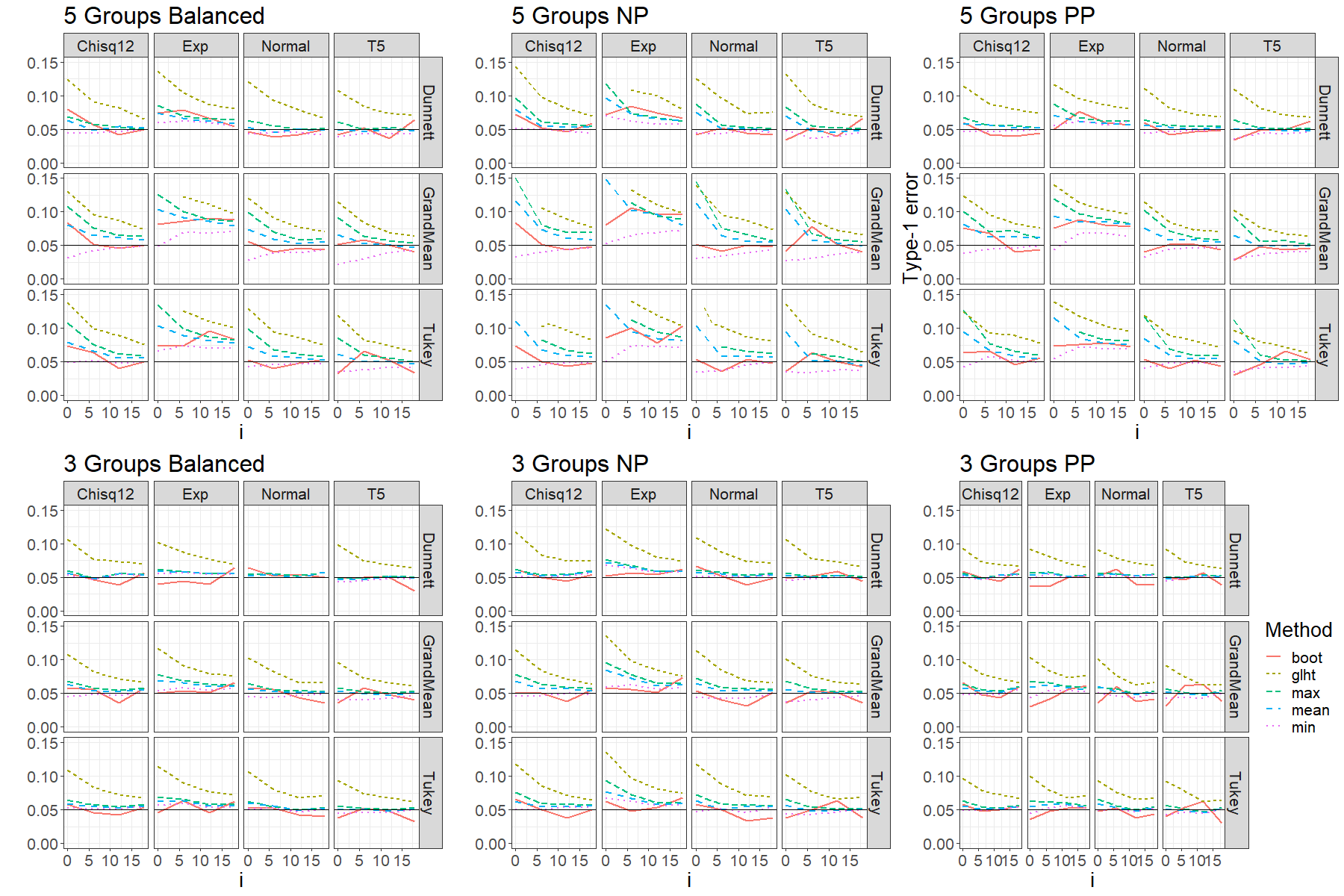}
\caption{Type-I error simulation balanced vs unbalanced with constant $N$}
\label{fig:constantN}
\end{figure}

To compare balanced and unbalanced designs with the same $N$ for both we also ran the following setting in addition to the ones discussed in \ref{sec4}. In the balanced case $n_j = 10 + i$ for all groups. In the unbalanced case $\mathbf{n} = (8, 10, 12) + i \text{ or } \mathbf{n} = (8,9,10,11,12) + i$ for 3 and 5 groups respectively with negative pairing and equivalently for positive pairing. Figure \ref{fig:constantN} shows the results.

\section{Negative Pairing vs Positive Pairing \label{app2}}
\begin{figure}[!htb]
\includegraphics[width = 140mm]{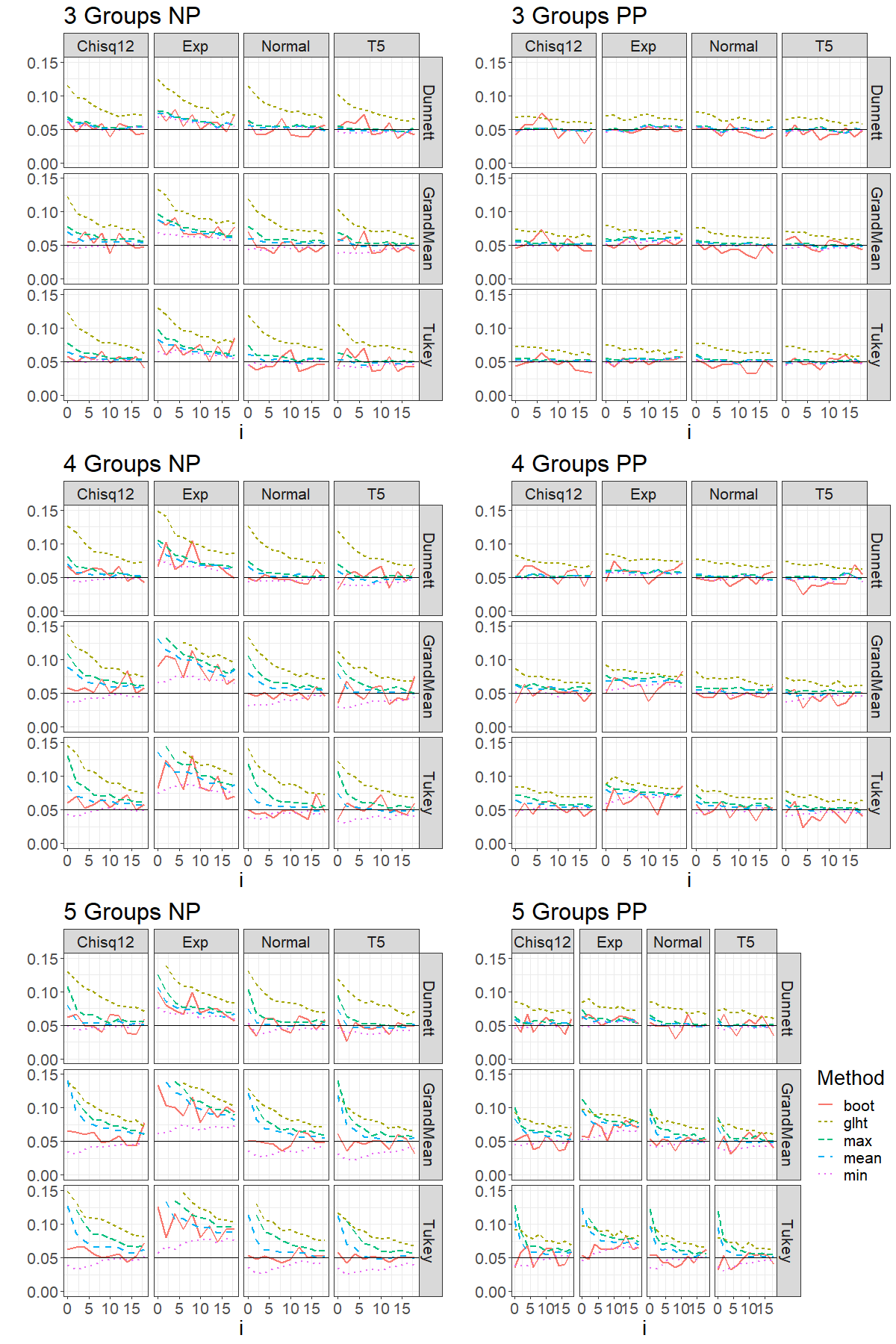}
\caption{Type-I error simulation results negative pairing vs positive pairing}
\label{fig:npvspp}
\end{figure}

\begin{figure}[!htb]
\includegraphics[width = 160mm]{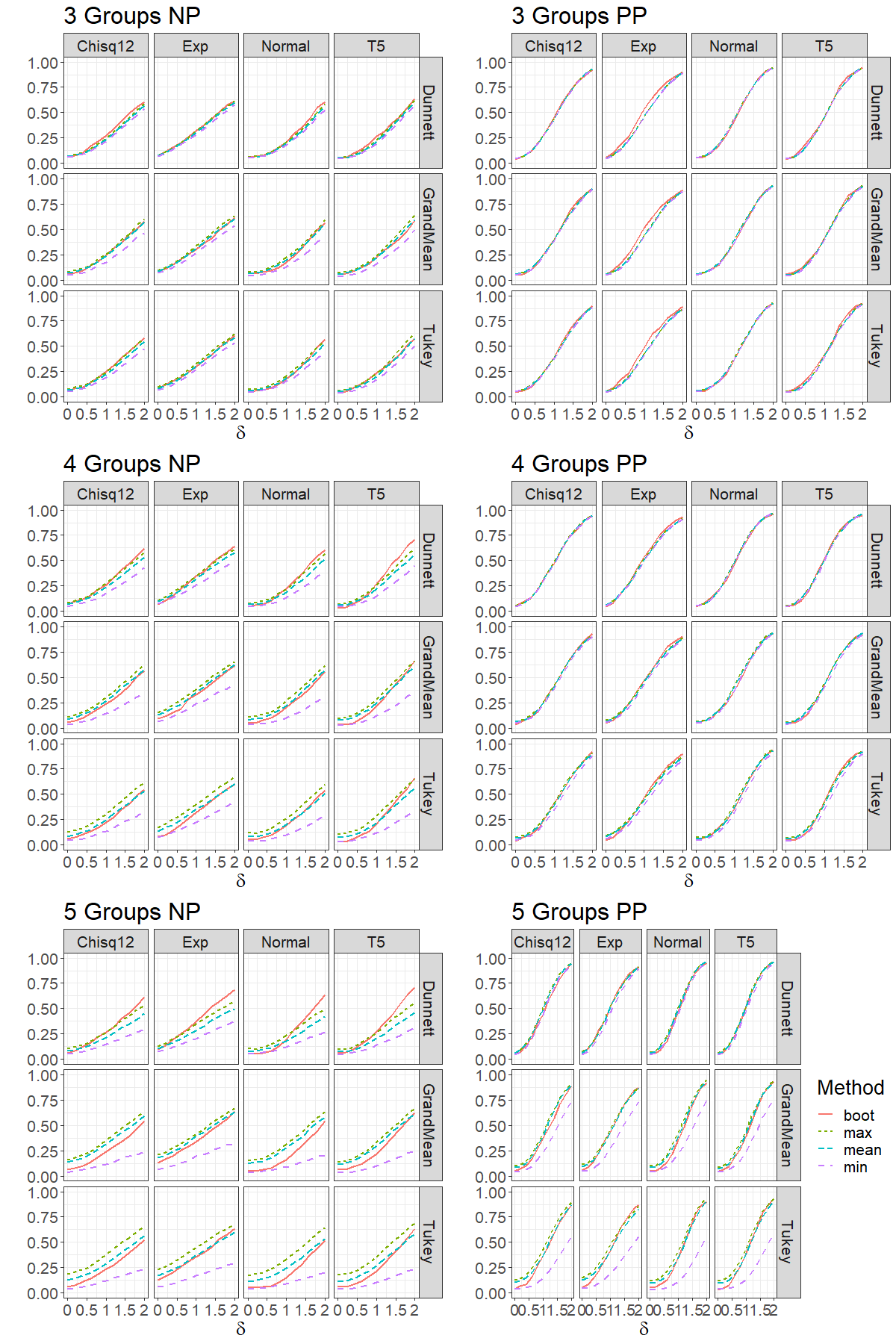}
\caption{Power simulation results negative pairing vs positive pairing}
\label{fig:npvspp_power}
\end{figure}

\section{Poweranalysis for Alternative 2 \label{app3}}
\begin{figure}[!htb]
\includegraphics[width = 160mm]{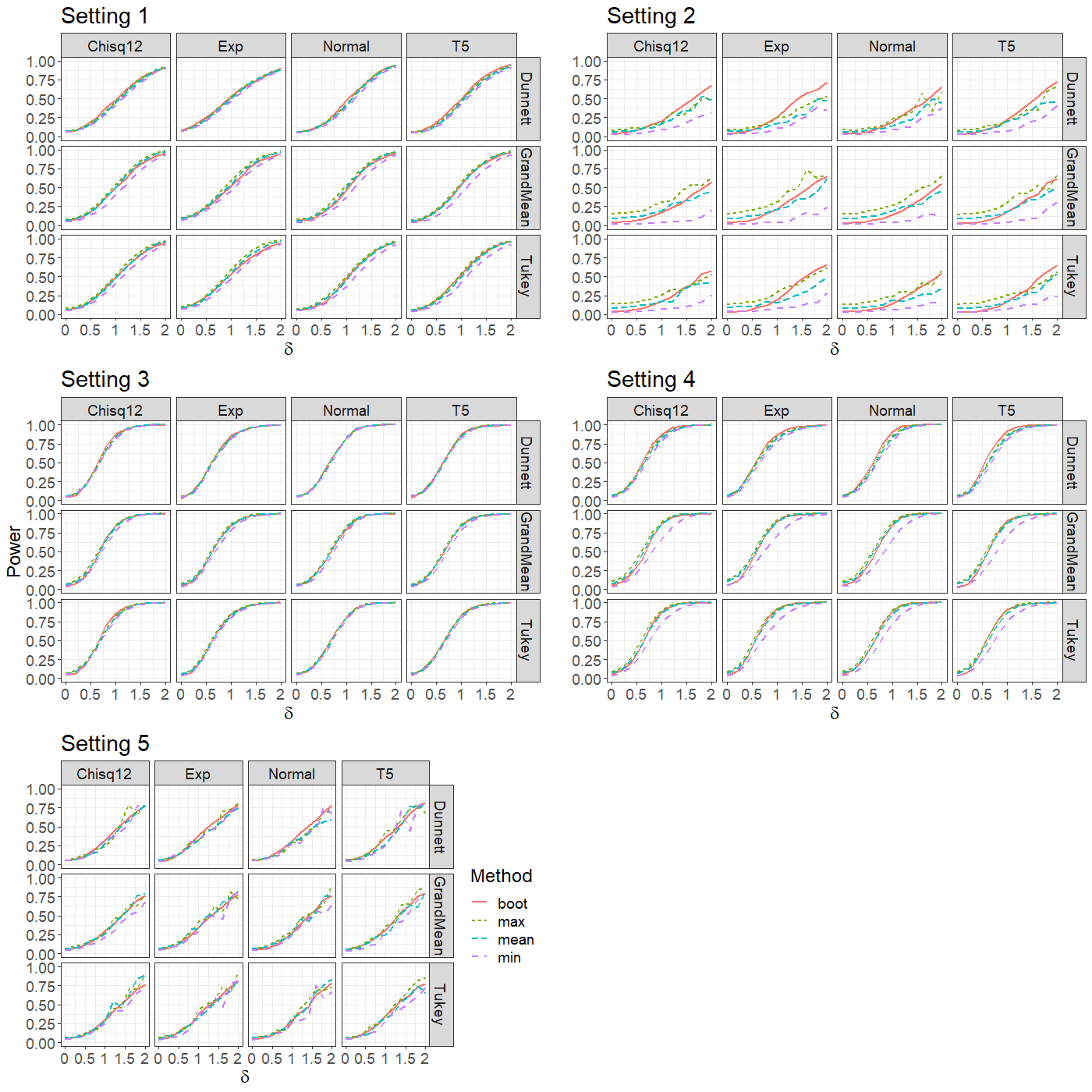}
\caption{Power simulation results for alternative 2}
\label{fig:power2}
\end{figure}
\end{document}